\documentclass[aps,prl,amsmath,amssymb,floatfix,twocolumn,amsmath,superscriptaddress,twocolumn,nofootinbib,tighten,letterpaper]{revtex4}
\usepackage{multirow}
\usepackage{bbold}
\usepackage{subfigure}
\usepackage{color}
\usepackage{mathrsfs}
\usepackage{hyperref}
\usepackage[normalem]{ulem}
\usepackage{bm}
\usepackage{pifont}

\renewcommand\vec[1]{\ensuremath\boldsymbol{#1}} 

\usepackage{amsfonts, relsize, color}
\usepackage{graphics}
\usepackage{graphicx}
\usepackage{subfigure}
\usepackage{hyperref}
\usepackage{color}
\usepackage{comment}

\renewcommand\vec[1]{\ensuremath\boldsymbol{#1}} 

\begin{document}
\title{Correlated fractional Dirac materials}

\author{Bitan Roy}\email{bitan.roy@lehigh.edu}
\affiliation{Department of Physics, Lehigh University, Bethlehem, Pennsylvania, 18015, USA}

\author{Vladimir Juri\v ci\' c}\email{vladimir.juricic@su.se}
\affiliation{Departamento de F\'isica, Universidad T\'ecnica Federico Santa Mar\'ia, Casilla 110, Valpara\'iso, Chile}
\affiliation{Nordita, KTH Royal Institute of Technology and Stockholm University, Hannes Alfv\'ens v\"ag 12, 106 91 Stockholm, Sweden}

\date{\today}
\begin{abstract}
Fractional Dirac materials (FDMs) feature a fractional energy-momentum relation $E(\vec{k}) \sim |\vec{k}|^{\alpha}$, where $\alpha \; (<1)$ is a real noninteger number, in contrast to that in conventional Dirac materials with $\alpha=1$. Here we analyze the effects of short- and long-range Coulomb repulsions in two- and three-dimensional FDMs. Only a strong short-range interaction causes nucleation of a correlated insulator that takes place through a quantum critical point. The universality class of the associated quantum phase transition is determined by the correlation length exponent $\nu^{-1}=d-\alpha$ and dynamic scaling exponent $z=\alpha$, set by the band curvature. On the other hand, the fractional dispersion is protected against long-range interaction due to its nonanalytic structure. Rather, a linear Dirac dispersion gets generated under coarse graining, and the associated Fermi velocity increases logarithmically in the infrared regime, thereby yielding a two-fluid system. Altogether, correlated FDMs unfold a rich landscape accommodating unconventional emergent many-body phenomena.
\end{abstract}

\maketitle

\emph{Introduction}.~Nodal Fermi liquids harbor a rich variety of electronic band dispersions around a few isolated points in the Brillouin zone that are often symmetry protected. They can display linear ($\alpha=1$), quadratic ($\alpha=2$) and cubic ($\alpha=3$) energy-momentum dispersion $E(k_j) \sim \pm |k_j|^\alpha$ along one or more than one component of the spatial momentum ($\vec{k}$), for example. Here $+$ ($-$) corresponds to the conduction (valence) band. Momentum is measured from the band touching point at $\vec{k}=0$. As such, isotropic but linear and nonlinear band dispersions can be observed in monolayer and multilayer graphene, respectively~\cite{graphene:RMP}, and in three-dimensional Dirac and Weyl semimetals~\cite{wsm:review}. Nodal Fermi liquids can also manifest mixed energy-momentum relations, where integer $\alpha$ is direction dependent, as in multi-Weyl semimetals~\cite{multiweyl:1, multiweyl:2, multiweyl:3,multiweyl:4} and at the quantum critical point (QCP) when the system resides at the brink of band insulation~\cite{SMInsQCP:1, SMInsQCP:2, SMInsQCP:3}. However, irrespective of microscopic details, the symmetry class and the dimensionality of the system, $\alpha$ is always an integer, hindering a large family of gapless electronic materials, where $\alpha$ can take real fractional values. We name them \emph{fractional Dirac materials} (FDMs), characterized by density of states (DOS) $\varrho \sim |E|^{d/\alpha-1}$ and frequency ($\Omega$) dependent diagonal optical conductivity (OC) $\sigma \sim \Omega^{(d-2)/\alpha}$ in the noninteracting limit.

\begin{figure}[t!]
\includegraphics[width=0.95\linewidth]{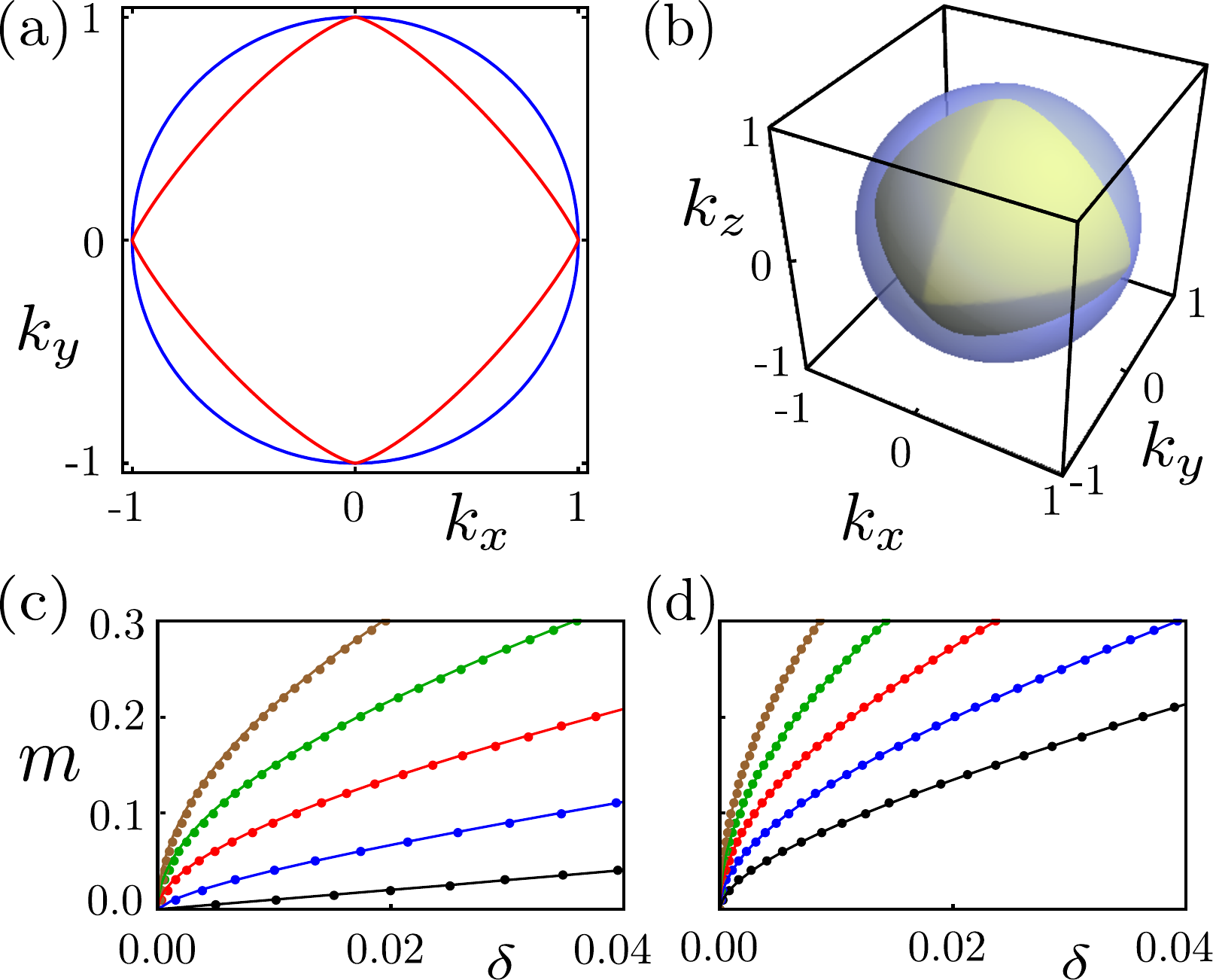}
\caption{Constant energy contours in (a) $d=2$ for $\alpha=1$ (blue) and $\alpha=2/3$ (red) (red) and (b) in $d=3$ for $\alpha=1$ (blue) and $\alpha=2/3$ (yellow). We set $v_\alpha=1$ and $\vec{k}$ is measured in units of $a^{-1}$, where $a$ is the lattice spacing [Eq.~\eqref{eq:hamilFDM}]. Self-consistent solutions of chiral symmetry breaking mass in (c) $d=2$ and (d) $d=3$ for $\alpha=1$ (black), $0.8$ (blue), $0.6$ (red), $0.4$ (green) and $0.2$ (brown), as a function of the reduced distance from a QCP ($\delta$). For $\alpha=1$ we recover a conventional Dirac system with linearly dispersing quasiparticles. Otherwise, the system describes an FDM of order $\alpha$. See text for details.
}~\label{Fig1}
\end{figure}

Here we investigate the effects of a short-range piece as well as of the long-range tail of the repulsive Coulomb interactions in two-dimensional (2D) and three-dimensional (3D) FDMs of order $\alpha <1$ for which the effective Hamiltonian scales as $|k_j|^\alpha$ [Eq.~(\ref{eq:hamilFDM}) and Fig.~\ref{Fig1}]. Due to the vanishing DOS in FDMs, sufficiently weak generic local or short-range interactions are irrelevant perturbations. However, beyond a critical threshold of interaction an FDM undergoes a continuous quantum phase transition (QPT) through a QCP and becomes a correlated Dirac insulator. We capture the university class of such a QPT, which is distinct from its counterpart in conventional Dirac systems ($\alpha=1$), within the framework of a Gross-Neveu model~\cite{gross-neveu}. The associated critical exponents in turn also govern the scaling of the spectral gap in the ordered phase with the interaction strength [Fig.~\ref{Fig1}]. We show that the QPT in 2D (3D) FDMs is non-Gaussian (mean-field or Gaussian) in nature.

In the presence of long-range Coulomb interaction, the electronic dispersion in 2D and 3D FDMs, on the other hand, remains invariant due to its nonanalytic structure [Eq.~(\ref{eq:hamilFDM})]. Rather under coarse graining, as the system approaches the deep infrared regime, a linear Dirac dispersion gets generated through the quantum many-body corrections. The effective Fermi velocity of such emergent conventional Dirac quasiparticles then increases logarithmically at lower energies. The resultant system thus describes a two-component quantum fluid at low energies, composed of (a) effectively noninteracting (protected by nonanalyticity of the dispersion) fractional and (b) marginally sharp (due to logarithmically increasing Fermi velocity) conventional Dirac fermions.

Altogether, correlated fractional Dirac liquids unfold a territory of rich emergent quantum phenomena, which can be studied using quantum Monte Carlo simulation of their lattice realizations in terms of infinitely long-ranged power-law hopping along the principal directions~\cite{syzranov:longrange} and possibly in electronic fractal lattices, recently realized on designer quantum materials~\cite{kempkes-fractal-exp, shang-fractal-exp}. Fractal lattices are promising platforms to host FDMs, where the characteristic irrational fractal dimension $d_{\rm frac}$ may give rise to noninteger $\alpha$.

\begin{figure}[t!]
\includegraphics[width=0.95\linewidth]{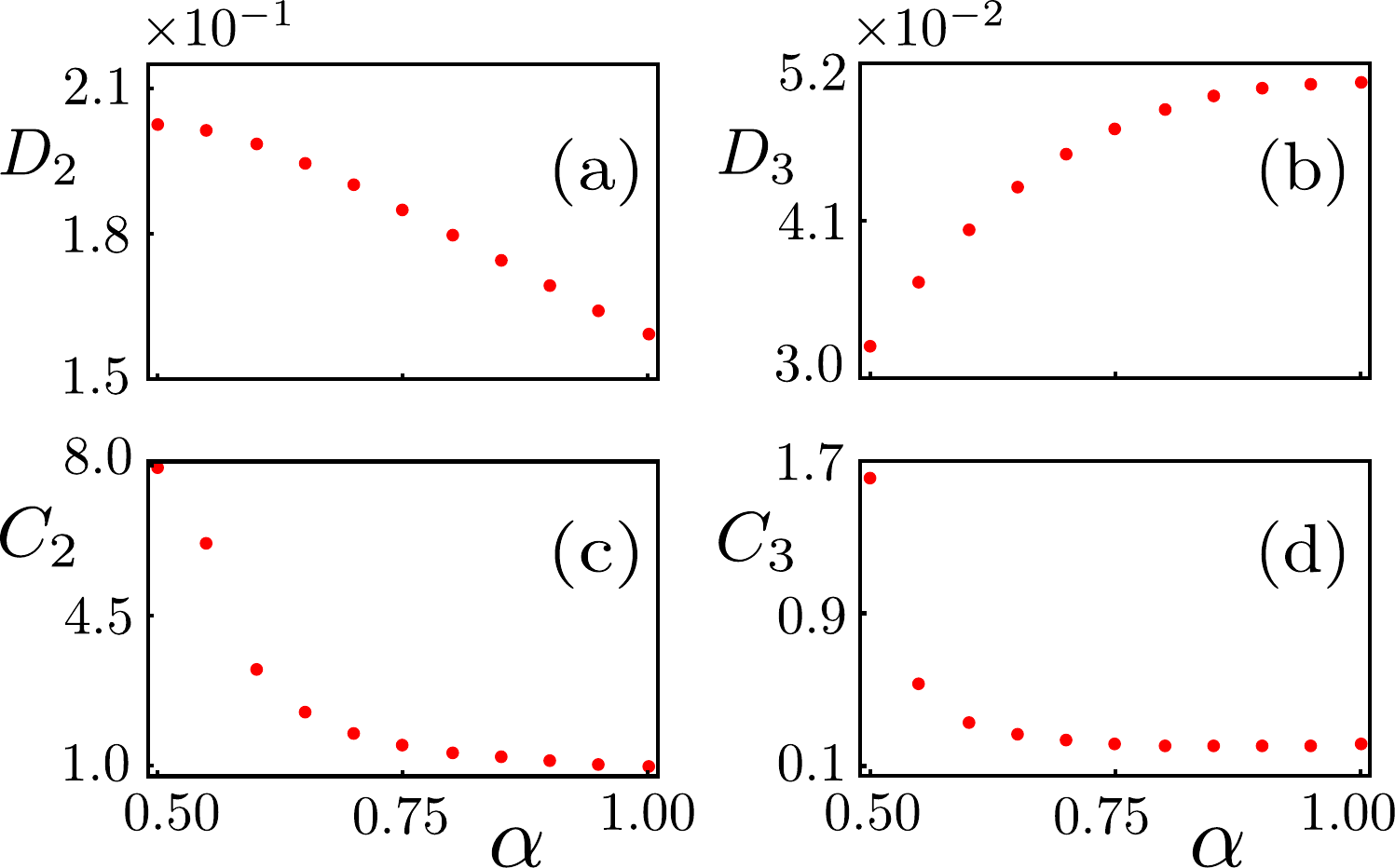}
\caption{Scaling of two universal functions $D_d$ [(a) and (b)] and $C_d$ [(c) and (d)], respectively governing the energy and frenqency dependances of DOS and OC in two ($d=2$) and three ($d=3$) dimensions. See Eq.~\eqref{eq:scalingDOSOC}.
}~\label{Fig2}
\end{figure}

\emph{Model}.~The effective Hamiltonian for a $d$-dimensional FDM of order $\alpha$ takes the form
\begin{equation}~\label{eq:hamilFDM}
H_{\rm FD} (\vec{k})= \sum^{d}_{j=1} v_\alpha |k_j|^\alpha \; \text{sgn}(k_j) \; \Gamma_j,
\end{equation}
where $v_\alpha$ bears the dimension of energy $\times \; a^{\alpha}$, such that $H_{\rm FD}(\vec{k})$ has the dimension of energy. We set the lattice spacing $a=1$. Mutually anticommuting $D$-dimensional Hermitian $\Gamma$ matrices, operating on the orbital or sublattice degrees of freedom, satisfy the algebra $\{ \Gamma_j, \Gamma_k \}=2 \delta_{jk} \mathrm{I}_D$, where $j,k=1, \cdots, D$ and $\mathrm{I}_D$ is a $D$-dimensional identity matrix. For now we keep $D$ arbitrary. The energy spectra of $H_{\rm FD}$ are $\pm E_{\alpha}(\vec{k})$, where
\begin{equation}
\hspace{-0.25cm}
E_{\alpha}(\vec{k})= v_\alpha \bigg[ \; \sum^{d}_{j=1} |k_j|^{2 \alpha} \; \bigg]^{\frac{1}{2}} \equiv v_\alpha |k|^\alpha \bigg[ \; \sum^{d}_{j=1} |\hat{\Omega}_j|^{2 \alpha} \; \bigg]^{\frac{1}{2}}
\end{equation}
and $\hat{\Omega}_j$ are the components of a $d$-dimensional spherical unit vector. Noninteracting $d$-dimensional FDMs are characterized by the power-law scalings of DOS and diagonal OC~\cite{supplementary}
\begin{eqnarray}~\label{eq:scalingDOSOC}
\varrho=D_d(\alpha) \; |E|^{\frac{d}{\alpha}-1}
\:\; \text{and} \:\;
\sigma= \frac{e^2}{h} \frac{\pi D}{16} C_d(\alpha) \left( \frac{\Omega}{v_\alpha} \right)^{\frac{d-2}{\alpha}},
\end{eqnarray}
respectively. The scalings of $D_d(\alpha)$ and $C_d(\alpha)$ are shown in Fig.~\ref{Fig2}. Thus specific heat ($C_v$) and compressibility ($\kappa$) in FDMs scales as $T^{d/\alpha}$ and $T^{d/\alpha-1}$, respectively, with temperature $T$.

The imaginary time ($\tau$) Euclidean action associated with $H_{\rm FD} (\vec{k})$ reads
\begin{equation}\label{eq:actionFDM}
S_0 = \int d\tau \int d^d\vec{x} \left\{ \Psi^\dagger \; \left[ \partial_\tau + H_{\rm FD}(\vec{k} \rightarrow -i \boldsymbol{\nabla}) \right] \; \Psi \right\},
\end{equation}
where we set $\hbar=1$, and $\Psi$ and $\Psi^\dagger$ are $D$-dimensional independent Grassmann variables. Under coarse graining $\tau \to e^{z \ell} \tau$ and $\vec{x} \to e^{\ell} \vec{x}$, where $\ell$ is the logarithm of the renormalization group (RG) scale and $z$ is the dynamic scaling exponent, measuring the relative scaling between the energy and momentum, with $z=\alpha$ in noninteracting FDMs. Then the scale invariance of the free action $S_0$ mandates the following scaling dimensions: $[\Psi]=[\Psi^\dagger]=d/2$ and $[v_\alpha]=z-\alpha$. Hence, in a noninteracting system $v_\alpha$ is scale invariant or marginal as $\alpha=z$ therein. Next we seek to scrutinize the effects of interactions among fractional Dirac quasiparticles. In what follows, we discuss the imprints of short-range and long-range Coulomb interactions on FDMs separately.

\emph{Short-range interaction}.~Short-range interactions among electronic quasiparticles in nodal Fermi liquids typically give birth to various spontaneously broken symmetry phases. Among them Dirac masses are most prominent at low temperatures. In the ordered phase, they give rise to isotropic gapped quasiparticle spectra, yielding electrical or thermal insulators, and lead to a maximal gain of the condensation energy. A Dirac mass is represented by the fermion bilinear $\Phi=\Psi^\dagger M \Psi$. The $D$-dimensional Hermitian matrix $M$ satisfies $\{M, \Gamma_j \}=0$ for $j=1, \cdots, d$ and $M^2= \mathrm{I}_D$. In the ordered phase, $\langle \Psi^\dagger M \Psi \rangle \neq 0$. In the spirit of the Gross-Neveu formalism, such a Dirac mass can be favored by a local or momentum independent four-fermion interaction for which the Euclidean action is~\cite{gross-neveu}
\begin{equation}
S^{\rm SR}_{\rm int} = \int d\tau \int d^d\vec{x} \; g_{_m} \left( \Psi^\dagger M \Psi \right)^2,
\end{equation}
where $g_{_m}$ is the corresponding coupling constant. For simplicity, here we consider a single-component microscopic Isinglike symmetry breaking scalar Dirac mass that yields an insulator in the ordered phase. The following discussion can be generalized to vectorlike Dirac masses, which, on the other hand, break a continuous symmetry following the spirit of the Nambu-Jona-Lasinio~\cite{NJL} model as well as for superconducting Dirac masses~\cite{roy-juricic:SC}. Leaving these cases for future investigations, here we focus on the nucleation of an Isinglike Dirac mass.

\begin{figure}[t!]
\includegraphics[width=0.95\linewidth]{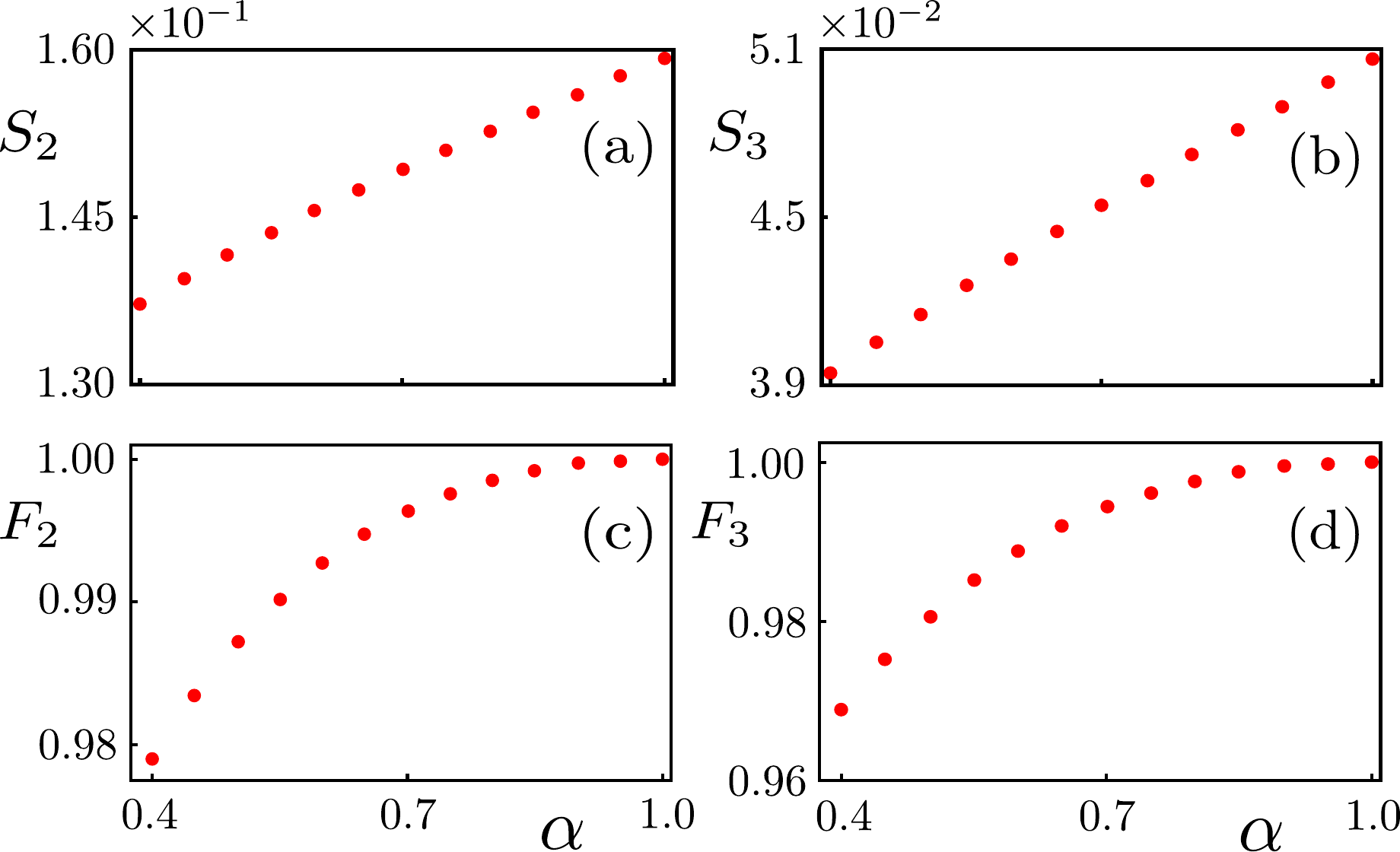}
\caption{Scaling of $S_d$ [(a) and (b)] and $F_d$ [(c) and (d)] respectively appearing in the definition of dimensionless Gross-Neveu interaction [Eq.~\eqref{eq:GRflowshortrange}] and the RG flow equation of Fermi velocity ($v_{_1}$) of emergent linearly dispersing Dirac quasiparticles due to long-range Coulomb repulsion [Eq.~\eqref{eq:flowv1}] in two ($d=2$) and three ($d=3$) dimensions.
}~\label{Fig3}
\end{figure}

The scale invariance of $S^{\rm SR}_{\rm int}$ leads to $[g_{_m}]=\alpha-d$. Therefore, when $\alpha<d$, yielding a vanishing DOS in FDMs, a sufficiently weak generic local quartic interaction is an \emph{irrelevant} parameter. Consequently, nucleation of any Dirac mass takes place beyond a critical threshold of interaction via a QPT occurring through a QCP located at $g_{_m}=g^\ast_{_m}$ (say). Here we capture such an emergent quantum critical phenomenon in a correlated FDM from a perturbative RG analysis controlled by a small parameter $\epsilon=\alpha-d$. Upon accounting for the leading order or one-loop quantum corrections we arrive at the RG flow equation for the dimensionless coupling constant $\lambda_m=g_{_m} \Lambda^{d-\alpha} S_d(\alpha)/v_\alpha$, explicitly given by~\cite{supplementary}
\begin{equation}~\label{eq:GRflowshortrange}
\beta_{\lambda_m}=\frac{d \lambda_m}{d \ell}=- \epsilon \lambda_m + (D-2) \lambda^2_m
\end{equation}
after integrating out fast Fourier modes with Matsubara frequencies $-\infty < \omega < \infty$ and residing within a thin Wilsonian momentum shell $\Lambda e^{-\ell} < |\vec{k}|<\Lambda$. Here $\Lambda$ is the ultraviolet momentum cutoff up to which the FDMs show fractional dispersion with power $\alpha$ [Eq.~(\ref{eq:hamilFDM})] and
\begin{equation}~\label{eq:sfunction}
S_d(\alpha)= \int \frac{d \hat{\Omega}}{(2 \pi)^d} \bigg[ \sum^d_{j=1} |\hat{\Omega}_j|^{2 \alpha} \bigg]^{-\frac{1}{2}} \equiv \int \frac{d \hat{\Omega}}{(2 \pi)^d} \frac{1}{[f(\hat{\Omega})]^{1/2}}.
\end{equation}
The scalings of $S_d(\alpha)$ in $d=2$ and $d=3$ are shown in Fig.~\ref{Fig3}. Notice that when the $\Gamma$ matrices are Pauli matrices ($D=2$), the perturbative corrections $\sim \lambda^2_m$ in Eq.~(\ref{eq:GRflowshortrange}) vanish. A similar conclusion holds for conventional Dirac fermions up to the two-loop order~\cite{gracey:twocomponent}, which, however, possibly undergo a continuous QPT at finite coupling~\cite{rosa-vitale-witerich}. Whether such a conclusion holds for FDMs remains to be investigated.

Thus, let us consider $D=4$, the minimal dimensionality of the $\Gamma$ matrices for which the quantum corrections in the RG flow equation of $\lambda_m$ [Eq.~(\ref{eq:GRflowshortrange})] are nontrivial. Then 2D fractional and conventional Dirac systems enjoy a continuous global SU(2) $\otimes$ U(1) chiral symmetry, generated by $\{\Gamma_{34},\Gamma_{45},\Gamma_{35} \}$ and $\Gamma_{12}$, respectively, where $\Gamma_{jk}=[\Gamma_j ,\Gamma_k]/(2 i)$. The Dirac mass breaks the continuous SU(2) chiral symmetry, unless $M=\Gamma_{12}$. This is so because the maximal number of mutually anticommuting four-component Hermitian matrices is five. By contrast, in $d=3$ the Dirac mass breaks a continuous U(1) chiral symmetry, generated by $\Gamma_{45}$.

The QCP describing the QPT between a nodal fractional Dirac liquid and a Dirac insulator is located at $\lambda_m=\lambda^\ast_m=\epsilon/(D-2)$. As the fermionic self-energy correction due to local interactions vanishes to the leading order in the $\epsilon$ expansion~\cite{supplementary}, this QCP is characterized by the correlation length exponent ($\nu$), given by
\begin{equation}
\nu^{-1}=\left.\frac{d\beta_{\lambda_m}}{d\lambda_m} \right\vert_{\lambda_m=\lambda^\ast_m}=\epsilon = d-\alpha, \:\:
\text{and} \:\:
z=\alpha.
\end{equation}
These two exponents determine the universality class of the fractional Dirac liquid to insulator QPT. In addition, they determine the scaling of the mass gap, which can be demonstrated by solving the self-consistent gap equation.

The self-consistent gap equation is obtained by performing a Hubbard-Stratonovich decomposition of the four-fermion term via a bosonic field $\Phi$ and subsequently integrating out the fermionic fields, yielding~\cite{supplementary}
\begin{equation}
\frac{1}{g_{_m}}= \int^\prime \frac{d^d\vec{k}}{(2 \pi)^d} \frac{1}{\left[ E^2_{\alpha}(\vec{k}) + \Phi^2 \right]^{1/2}} \equiv F(\Phi).
\end{equation}
The momentum integral is restricted up to the ultraviolet cutoff $\Lambda$ (denoted by the `prime' symbol). The right-hand side of this gap equation scales as $\Lambda^{d-\alpha}$ and it is ultraviolet divergent for $d>\alpha$. Such an ultraviolet divergence can be regularized by defining a critical coupling for the ordering $g^\ast_{_m}=[F(0)]^{-1}$, in terms of which we arrive at the regularized gap equation
\begin{equation}~\label{eq:gapfunction}
\hspace{-0.25cm}\delta= \int^1_0 x^{d-1} \left[ \frac{1}{x^\alpha} -\int \frac{d\hat{\Omega}}{(2 \pi)^d} \frac{ [S_d(\alpha)]^{-1}}{[x^{2 \alpha} f(\hat{\Omega})+ m^2]^{1/2}}  \right] dx,
\end{equation}
where $x=k/\Lambda$ and $m=\Phi/(v_\alpha \Lambda^\alpha)$ are dimensionless, and the reduced distance from the QCP is $\delta=(\lambda_m-\lambda^\ast_m)/(\lambda_m \lambda^\ast_m)$. We numerically solve this gap equation in $d=2$ and $d=3$. The results are shown in Fig.~\ref{Fig1}.

Notice that a nontrivial solution for the mass gap $m$, and concomitantly for the order parameter, exists only when $\delta>0$ or $\lambda_m>\lambda^\ast_m$, i.e., when interaction strength $\lambda_m$ is above a critical one ($\lambda^\ast_m$). The scaling of $m$ with $\delta$ in FDMs is distinct from its counterparts in conventional Dirac systems, where the mass gap shows linear (as $\nu=z=1$) and square-root (as $2\nu=z=1$) scaling with $\delta$ in $d=2$ and $d=3$, respectively, since $m \sim \delta^{\nu z}$. In a 3D Dirac system the scaling of $m$ with $\delta$ displays a logarithmic correction due to the violation of the hyperscaling hypothesis, as the system then lives at the upper critical dimension $d_{\rm up}=3$~\cite{roydassarma}. The upper critical dimension for FDMs is $d_{\rm up}=2 + \alpha$, where $\nu=1/2$. Therefore, 2D FDMs always remain below the upper critical dimension and the QPT is non-Gaussian in nature. By contrast, a 3D FDM always lives above the upper critical dimension, and the QPT is Gaussian or mean-field in nature.

\emph{Long-range Coulomb interaction}.~The instantaneous long-range Coulomb interaction is known to renormalize the Fermi velocity of linearly dispersing Dirac quasiparticles in both $d=2$~\cite{gonzalez-NPB} and $d=3$~\cite{goswami-PRL, hosur-PRL, roy-juricic:longrange3D}. This interaction is represented by a scalar gauge field $a_0(\tau,\vec{x})$ minimally coupled to the density of the quasiparticles. The corresponding propagator exhibits a dimensionality-dependent scaling with momentum $\sim |\vec{k}|^{d-1}$, ensuring the characteristic $1/r$ behavior of the density-density Coulomb interaction in real space in any $d$. The form of the Coulomb propagator, in turn, dictates that the electric charge ($e_d$) can (cannot) receive perturbative corrections in $d=3$ ($d=2$). The Coulomb  part of the Euclidean action therefore reads as
\begin{equation}
S_{\rm C} = \int d\tau \int d^d\vec{x}   \left(-ie_d a_0 \Psi^\dagger  \Psi +a_0\frac{1}{2|{\boldsymbol \nabla}|^{d-1}}a_0 \right).
\end{equation}

To analyze the effect of the long-range Coulomb interaction on fractional Dirac fermions, we compute the one-loop self-energy diagram. We find that the parameter $v_\alpha$ appearing in Eq.~\eqref{eq:hamilFDM} does not renormalize when $\alpha\neq1$, as a consequence of the nonanalytic structure of the dispersion of fractional Dirac excitations. However, the long-range Coulomb interaction generates linearly dispersing quasiparticles, for which the Hamiltonian is given by Eq.~\eqref{eq:hamilFDM} with $\alpha=1$. The RG flow equation for the Fermi velocity ($v_{_1}$) of emergent linearly dispersing quasiparticles in $d=2$ and $d=3$ reads~\cite{supplementary}
\begin{equation}\label{eq:flowv1}
\frac{d v_{_1}}{d\ell}= \frac{1}{C_d}\, F_d(\alpha) \, \alpha_{_{\rm FS}} \, v_{_1},
\end{equation}
where $C_2=8\pi$, $C_3=6\pi^2$, and the function (Fig.~\ref{Fig3})
\begin{equation}\label{eq:Fd-def}
F_d(\alpha)=\alpha\int \frac{d \hat{\Omega}}{N_d} \; \frac{|\hat{\Omega}_i|^{\alpha-1}}{[f(\hat{\Omega})]^{1/2}} \; \bigg(1-\frac{|\hat{\Omega}_i|^{2\alpha}}{f(\hat{\Omega})}\bigg),
\end{equation}
with $N_2=\pi$ and $N_3=8\pi/3$. The associated effective fine structure constant is $\alpha_{_{\rm FS}}={e_d^2}/{v_{_1}}$. The function $f(\hat{\Omega})$ is defined in Eq.~\eqref{eq:sfunction}, and  $F_d(\alpha)$ is independent of the choice of the component ($i=1, \cdots, d$) of the $d$-dimensional unit vector ($\hat{\Omega}_i$) at least in $d=2$ and $3$.

Notice that the Fermi velocity $v_{_1}$ grows logarithmically as the system approaches the deep infrared regime under coarse graining, as $F_2(\alpha)$ and $F_3(\alpha)$ are both positive definite. Furthermore, as the power of the fractional dispersion increases for $0<\alpha\leq1$, both functions increase monotonically, as shown in Fig.~\ref{Fig3}. This is, indeed, consistent with the scaling of the DOS with energy. Namely, as the DOS increases with increasing $\alpha$, the flow of the Fermi velocity of generated Dirac quasiparticles also increases. When the bare quasiparticles are linearly dispersing, well-known results for the flow of the Fermi velocities in $d=2$ and $d=3$ are readily recovered~\cite{gonzalez-NPB, goswami-PRL, hosur-PRL, roy-juricic:longrange3D}. Thus, long range Coulomb interaction in FDMs gives birth to a two-component quantum fluid constituted by (a) effectively noninteracting fractional and (b) marginal (due to logarithmically increasing Fermi velocity) conventional Dirac quasiparticles.

Finally, we show that  in $d=3$ the long-range tail of the Coulomb interaction is screened by fractional Dirac excitations. To this end,  we compute the polarization (bubble) diagram, and find that the Coulomb charge in $d=3$ is logarithmically decreasing under coarse graining, with the RG flow equation of the form
\begin{equation}\label{eq:RGflow-charge}
\frac{de^2_d}{d\ell}=-\alpha_{\rm FS}^{(\alpha)}\,e^2_d\,\mathcal{F}(\alpha) \:\:
\text{or} \:\:
\frac{d \alpha_{\rm FS}^{(\alpha)}}{d\ell}=-\left[\alpha_{\rm FS}^{(\alpha)} \right]^2 \,\mathcal{F}(\alpha).
\end{equation}
Here $\alpha_{\rm FS}^{(\alpha)}={e^2_d}/(v_\alpha\Lambda^{\alpha-1})$ is the effective dimensionless fine structure constant of the FDM and the function
\begin{equation}
\mathcal{F}(\alpha)=\frac{\alpha}{4}\big[(1-3\alpha) I(\alpha)+2\alpha J(\alpha)\big],
\end{equation}
for $1/2<\alpha\leq1$, with $I(\alpha)$ and $J(\alpha)$ defined in terms of the components of the three-dimensional unit vector
\begin{align}
I_{\rho\sigma}(\alpha)&=\int\frac{d\hat{\Omega}}{(2\pi)^3}\; \frac{|\hat{\Omega}_\rho|^{2\alpha-2}|\hat{\Omega}_\sigma|^{2\alpha-2}\hat{\Omega}_\rho \hat{\Omega}_\sigma}{[f(\hat{\Omega})]^{5/2}}\equiv I(\alpha)\delta_{\rho\sigma},\nonumber\\
J_{\rho}(\alpha)&= \int\frac{d\hat{\Omega}}{(2\pi)^3}\; \frac{|\hat{\Omega}_\rho|^{2\alpha-2}}{[f(\hat{\Omega})]^{3/2}}\equiv J(\alpha).
\end{align}
The function $\mathcal{F}(\alpha)$ is strictly positive, implying that the charge $e_d$ and fine structure constant $\alpha_{\rm FS}^{(\alpha)}$ of the FDM decrease during the RG flow. Furthermore, the fine structure constant of generated Dirac quasiparticles $\alpha_{_{\rm FS}}$ also decreases but even faster than in a conventional relativistic Dirac material due to additional screening by microscopic fractional Dirac excitations.

\emph{Summary and discussion}.~Here we explore the emergent quantum critical behavior of correlated fractional Dirac liquids in two and three dimensions in the presence of short-range and long-range Coulomb interactions. Strong short-range interactions can give rise to spontaneous symmetry breaking. By contrast, the long-range component of Coulomb interaction gives birth to linear Dirac dispersion through a spectral reconstruction, while leaving the original fractional Dirac dispersion unaffected, thereby yielding a two-component quantum fluid in the infrared regime. In future, we will develop a quantum critical description for the spontaneous symmetry breaking in terms of strongly coupled boson-fermion Mott-Yukawa theory~\cite{zinnjustin:book}, which will allow us to capture hallmarks of the emergent non-Fermi liquid near the associated QCP and the role of retarded boson-fermion Yukawa interaction in possible restoration of the Lorentz symmetry in correlated FDMs~\cite{RJH:JHEP}. Furthermore, by incorporating the retarded current-current interaction, accompanying the instantaneous density-density Coulomb repulsion, we will develop quantum electrodynamics for FDMs.

Our field-theoretic predictions on spontaneous symmetry breaking and associated quantum critical phenomena in FDMs subject to short-range Coulomb or Hubbardlike interactions can be tested, for example, from quantum Monte Carlo simulations of the lattice regularized tight-binding models for FDMs in terms of infinitely long-ranged power-law hopping~\cite{supplementary}, given that such a study exists for similar models of conventional Dirac fermions with linear dispersion, subject to the on-site Hubbard repulsion~\cite{QMC:SLACHubbard}. Effects of long-range Coulomb interactions and the resulting two-component quantum fluid can also be demonstrated using quantum Monte Carlo simulations as similar computations have already been performed on graphene's honeycomb lattice harboring conventional Dirac quasiparticles~\cite{QMC:CoulombGraphene}. Besides designer fractal materials~\cite{kempkes-fractal-exp, shang-fractal-exp}, quantum circuits constituted by arrays of superconducting qubits stand as a promising platform for the realization of quantum FDMs and proposed quantum phenomena therein, where the Sachdev-Ye-Kitaev model~\cite{Quantumcircuit:SYK} and a number of quantum hyperbolic lattices~\cite{Quantumcircuit:Hyperbolic} have already been engineered to showcase their exotic quantum many-body phenomena in tabletop experiments.

\emph{Acknowledgments}.~B.R. was supported by NSF CAREER Grant No.\ DMR- 2238679. V.J. acknowledges support from the Swedish Research Council (VR 2019-04735) and Fondecyt (Chile), Grant No. 1230933. Nordita is partially supported by Nordforsk.

\end{document}